\documentclass[12pt,fleqn]{article}
\usepackage[latin1]{inputenc}    
\usepackage[intlimits]{amsmath}
\usepackage{amssymb}
\usepackage{graphicx}
\usepackage{epsfig}
\begin{document}
\baselineskip 18pt  
\title{OZI violation in low energy $\omega $ and $\phi $ production in the pp system 
in a quark-gluon model \thanks{supported in part by the Forschungszentrum FZ Jülich (COSY)}
\thanks{preprint FAU-TP3-06/Nr. 10}}

\author{M. Dillig \thanks{email: mdillig@theorie3.physik.uni-erlangen.de} \\
Institute for Theoretical Physics III \\ 
University Erlangen-Nürnberg,  \\ Staudtstr. 7, D-91058 Erlangen, Germany}
\date{}
\begin{titlepage}
\maketitle
\begin{abstract}
We investigate OZI violation in near-threshold $\omega$ and $\phi $ production in the $pp$-system. 
Assuming ideal $\omega /\phi $ mixing (corrections are estimated), the energy dependence of the ratio
$R_{\omega / \phi} $ is analyzed in a perturbative quark-gluon exchange model up to the third other in the 
strong coupling constant $\alpha_{s}$ with the proton represented as a quark - scalar diquark system. 
We give a very natural explanation of the violation of the OZI rule in $\omega /\phi $ production 
and its energy dependence near the production thresholds.
\end{abstract}
\end{titlepage}
\newpage
\setcounter{page}{2}
The OZI rule, formulated by Okubo, Zweig and Iizuka \cite{Oku} - \cite{Iiz} provides a direct link to the 
quark-structure of hadrons in QCD: it states that meson production via disconnected quark lines is 
suppressed relative to connect $q \bar{q}$ excitations. One interesting example for this conjecture is the 
production of $\omega$ and $\phi $ mesons. Assuming ideal SU(3) mixing of octet-singlet representation, 
then 
\begin{displaymath}
| \omega \rangle = \cos \theta_{v} | \eta _{8} \rangle + \sin \theta_{v} | \eta _{1} \rangle 
= \frac{1}{\sqrt{2}} \; (u \bar{u} + d \bar{d}) 
\end{displaymath}
\begin{displaymath}
| \phi \rangle = - \sin \theta_{v} | \eta_{8} \rangle + \cos \theta_{v} | \eta_{1} \rangle 
= s \bar{s} \, ,
\end{displaymath}
i.~e. the $\phi$-meson is a pure $s \bar{s}$ state and its excitation is OZI forbidden. Deviations from 
ideal mixing \cite{eidel} yield the ratio \cite{lipk}
\begin{displaymath}
R_{OZI} = \frac{\sigma (i \to f \phi)}{\sigma (i \to f \omega)} = tg^{2} (\Delta \theta_{v} 
= 3.7^{0}) = 4.2 \cdot 10^{-3}\,.
\end{displaymath}
$R_{\phi /\omega} $ has been investigated extensively for various systems, such as in $\pi N $
\cite{titov, sibir}, $\gamma N $ \cite{sibmeis, tit03}, (radiative decays of) vector mesons 
\cite{gloz, meiss}, $N \bar{N} $ annihilation \cite{filippi, marku} and $NN $ systems \cite{nomok,nakay}.
In this note we focus on near-thereshold $\phi, \omega $ production in pp collisions,
where recently data on the total $\omega, \phi $ cross sections and on the $ \phi/\omega $ ratio 
have become available  \cite{disto}- \cite{hibou}.  
\vskip 0.3cm

Currently, most investigations on near threshold $pp \to pp \omega/\phi $ production are based on
meson exchange models both for $\omega $ \cite{nakay} -\cite{sibi96} and $\phi $ production 
\cite{sibir55} - \cite{haid}, with or without the inclusion of baryon resonances \cite{tsush, fuchs}.
As the $\omega$ and the $\phi $ meson carry the same quantum numbers $(J^{\pi} = 1^{-}, T = 0 )$, 
the leading contributions (without baryon resonances) involve the  $(\omega,\phi)  \rho \pi $ 
and $(\omega, \phi) pp $  coupling (Fig. 1). Now $g_{\omega \rho \pi} $ and $g_{\phi \rho \pi}$ can be extracted,
via vector meson dominance, from their decay into the $\gamma \pi$ channel \cite{meiss}, while
$g_{\omega pp} $ is  controlled from modern meson exchange models for the  NN interaction \cite{machl}.
Thus a comparison of $\phi/\omega $ production as a function of the excess energy 
$Q = \sqrt{s} - (2 M_{p} + m_{\phi,\omega}) $ should provide detailed information on $g_{\phi pp} $, which 
involves a typical uncertainty up to one order of magnitude in the literature \cite{tsush}.
Exploring this uncertainty, meson exchange models give qualitatively the right trend: an increasing OZI suppression
of $\phi $-production for decreasing Q (in comparing with the data, the $\phi/\omega $ ratio is normalized 
to the experimental ratio at the DISTO energy \cite{disto}, as the poor knowledge of the pp initial state 
interaction at the relevant energies prohibits a quantitative normalization of the $\phi, \omega $ cross sections
at momentum transfers of typically 1 GeV/c). 
\vskip 0.3cm

Opposite to meson-exchange models we follow the quark-based OZI arguments more explicitly and construct the 
contributions to $\phi /\omega $ production in a quark-gluon model . The leading terms up to third order
gluon exchange as summarized in Fig. 2. Both sets of diagrams in 2 (a,b) favor $\omega $ versus $\phi $ 
production. For the 'mesonic' component the basic difference stems from the 2. diagram in fig. 2 (a). 
For ideal $\phi /\omega $ mixing, only the $\omega$ can be produced by the exchange of two gluons (without or with interchange of two quark lines), whereas $\Phi$ production is strictly forbidden. For the 'nucleonic' component 2 (b) the difference is similarly striking: the leading 3g-exchange piece in $\omega$-production, i.~e. colorless (Pomeron) 
2g exchange, followed by $q \bar{q} $ excitation, is again strictly forbidden for $\phi $ production, where the direct excitation of the $s \overline s$ quark-anti quark in the $\Phi$ meson without the interchange of quark lines yields the only contribution. Collecting just for a very qualitative estimate the corresponding color matrix elements (and adding thereby noncrossed and crossed gluon diagrams and leaving out common factors)
\begin{displaymath}
R^{1}_{\text{colour}} = \frac{|M^{no \,ex}_{3g} |^{\phi}_{2}}{|
M^{no\, ex}_{2g} + M^{ex}_{2g}
+ M^{no\, ex}_{3g} + M^{ex}_{pom} |^{2}_{\omega}} =
\left (\frac{7}{7 + 8 + 32} \right )^{2} = 2.2 \cdot 10^{-2}
\end{displaymath}
$\phi $ production is suppressed by roughly 2 orders of magnitude relative to $\omega $ production.
\vskip 0.3cm
The main steps and approximations for a detailed calculation are readily summarized. As energy 
and momentum transfers $\Delta E \sim \Delta q \sim 1$ GeV are far below the onset of perturbative  QCD,
we model the transition amplitude with effective quark and gluon degrees of freedom. The transition 
operator, integrated over the internal proton and meson Jacobi-coordinates (in coordinate space) 
is given as $( \lambda = \phi, \omega) $
\begin{displaymath} 
M_{qq \to qq (q \bar{q})} (\underline{R})  =   \langle \phi_{p} (r,\rho) \phi_{p} (r^{\prime}, \rho^{\prime}) 
\phi_{\lambda} (r_{\lambda}) |   V_{q \to q(q \bar{q}} (\underline{R}) V_{qq - qq} (\underline{R}) |   
 | \phi_{p} (r, \rho) \phi_{p} (r^{\prime}, \rho^{\prime}) \rangle
\end{displaymath} 
as a function of the relative pp coordinate $\underline{R} $. Above the $qq$ and $q \to q (q\bar{q}) $
interaction is derived from the relativistic one-gluon exchange operator \cite{furu}, \cite{ebert}, where the corresponding
(particle, antiparticle) Dirac spinors are expanded up to 
$ \left ( \frac{q}{\omega(q) + m_{q}} \right )^{2} $
in the quark mass $m_{q} = 330 $ MeV and the quark energy $\omega (q) $, respectively (note that for 
equal momentum sharing among the quarks of the protons with $q \sim p /3$ ($p \sim $ 1 GeV/c 
is the proton momentum in the initial state), the expansion parameter
yields $\sim (2/5)^{2} $). Then \cite{muller} - \cite{fuji}
\begin{displaymath}
V_{qq - qq} (\underline{r}_{ij}) = \frac{4 \pi \alpha_{s}}{m_{g}^{2}} \; 
\frac{\underline{\lambda}_{i} \underline{\lambda}_{j}}{4} \; 
( V_{c} + V_{ss} + V_{LS} + V_{T}) e^{- \frac{mg^{2}}{4} \underline{r}^{2}_{ij}}
\end{displaymath}
with the strong coupling constant $\alpha _{s} \sim 2 $, the constituent gluon mass $m_{g} \sim 800$ MeV 
\cite{giacosa, bowman} and the central, spin, spin-orbit and tensor components; $\underline{\lambda}_{i,j}$  are SU(3) 
color matrices. Similarly we obtain for the $q \rightarrow q (q \overline q)$ excitation
\begin{displaymath}
V_{q-q(q \bar{q})} (\underline{r}_{ij}) = -  \frac{\alpha s}{8 \sqrt{\pi}} \; \frac{m^{3}_{g}}{m_{q}} \; 
\frac{\underline{\lambda}_{1} \underline{\lambda}_{2}}{4} \; 
\left ( (\underline{\sigma}_{i} x \underline{\sigma}_{j}) \underline{r}_{ij} \;
e^{- \frac{m^{2}_{g} \underline{r}^{2}_{ij}}{4}} +  i \; \frac{4}{m^{2}_{g}} \; 
e^{- \frac{m^{2}_{g} \underline{r}^{2}_{ij}}{4}}\; \underline{\sigma}_{j} 
\underline{\nabla}_{i} \right )
\end{displaymath}
(In practice the radial interactions are expanded as a superposition of Gaussians with different strength 
and width parameters). With these ingredients the 'Pomeron' exchange is formulated from the noncrossed
and crossed 2 g exchange, coupled to a color singlet \cite{donna} - \cite{lands} state
\vskip 0.3cm
Similarly, the proton and the mesons are also modelled in a Gaussian basis. Thus the vector meson 
wave functions are represented as 
\begin{displaymath}
\phi_{\lambda} (r_{ij}) = e^{- \frac{r^{2}_{ij}}{2 a_{\lambda}^{2}}} \; 
\Big [1/2 (i) 1/2 (j) \Big ] ^{1m,00}_{\text{spin, flavour}} \; 
\Big ( \frac{\delta_{ij}}{\sqrt{3}} \Big ) _{\text{colour}}
\end{displaymath}
with $a_{\lambda} = \sqrt{\frac{2}{3}} \; r_{rms} $ (i.~e. the root mean square radius of the meson). For the
proton we introduce an additional approximation to simplify the complicated $6q $ and $6q(q\overline{q}))$
many body problem: we represent the 
proton as a quark-scalar  diquark system
\begin{displaymath}
\phi_{p} (r, \rho) _{ijk} = \sum^{3}_{n = 1} c_{n} \; e^{- \frac{r^{2}+\rho^{2}}{6 a _{n}}}  
\Big [1/2 (i) [1/2 (j) 1/2(k)]^{0,0} \Big ]^{1/2 \;\mu_{p}, 1/2\; 1/2}_{\text{spin, flavour}} \cdot
\frac{\epsilon_{ijk}}{\sqrt{6}}
\end{displaymath}
with the parameters $c_{n}, a_{n} $ from resonating group calculations of the proton 
\cite{muller}- \cite{burger}. 
Treating the scalar diquark as a boson with mass $m_s$ (without antisymmetrizing its quark structure with the 
additional quark) dramatically simplifies antisymmetrization and the calculation of the spin-flavor-color
matrix elements. We remark that the quark-scalar diquark  configuration of the proton is 
supported from strong $qq $ correlations in scalar diquarks with $S = T = 0$ \cite{gockeler} - \cite{anselmino};
opposite,  the probability of axial diquarks with $S = T = 1 $ is suppressed by more than one order of 
magnitude compared to scalar diquarks in the proton (\cite{mineo}). 
\vskip 0.3cm
In a final step we calculate the total $\omega, \phi $ cross sections and their ratio as a function of Q
\begin{displaymath}
\sigma (Q) \sim \int \sum_{\text{spins}} | M_{pp \to pp \lambda} (Q) |^{2} \frac{1}{2 \omega_{\lambda}(k)} \; 
\delta (\underline{p}^{\prime}_{1} + \underline{p}^{\prime}_{2} + \underline{k})\; 
 \delta \; (\sqrt{s} - E_{p^{\prime}_{1}} - E_{p^{\prime}_{2}} 
- \omega_{\lambda} (k)) d V_{ps}
\end{displaymath}
($dV_{ps} $ denotes the integration over the 3-body phase space) with 
\begin{displaymath}
M_{pp \to pp \lambda} (Q) = \langle \chi^{f}_{pp} (\underline{R}) | M_{qq \to qq[q \bar{q})} (\underline{R}) | 
\chi^{i}_{pp} (\underline{R}) \rangle \,.
\end{displaymath}
Neglecting the meson-proton interaction, $\chi^{f}_{pp} (\underline{R}) $ is the distorted $pp $ wave, 
which is obtained for the $pp $ final state interaction from an expansion in a scattering length-effective range 
parametrization (\cite{kleef}). Without quantitative information on the $pp $ initial state interaction 
we use 
\begin{displaymath}
| \chi^{i}_{pp} (\underline{R}) \rangle = e^{\underline{i p R}} \cdot 
\sqrt{\frac{\sigma^{e\ell}_{pp}}{\sigma^{\text{total}}_{pp}}} 
\sim \frac{e^{\underline{ipR}}}{2}
\end{displaymath}
(with $\pm \, \underline{p}$ being the CM momentum of the protons in the initial state; the pp cross 
sections are taken from ref \cite{eidel}); other estimates from the literature \cite{hanhart}
yield a qualitatively similar result. 
\vskip 0.3cm

The results of our calculation are presented in Figs.~3 to 5. Figs. 3,4 show the Q-dependence of the $\omega $ and the $\phi$ cross section. Opposite to $\phi $-production, where only one data point  
is published together with still preliminary data from ANKE \cite{hartm}, 
a more detailed comparison is possible for the $\omega $ meson.
Within the given parameters  $(\alpha_{s} = 1.7, m_{s} = 600 \; \text{MeV}, m_{g} = 800 \; \text{MeV})$
we reproduce qualitatively the experimental $Q $-dependence. There remains still a significant 
parameter sensitivity, especially on final state $pp $ interactions and the $q $-diquark structure of the 
proton. For $\phi $-production, which is calculated with the parameters fixed from $\omega $ production,
the $Q $-dependence is similar, though the relative strength of 
$\sigma_{pp \to pp \phi} (Q)/\sigma_{pp \to pp \phi} (Q_{0} = 83 $ MeV) with $Q < Q_{0} $ decreases 
relative to $ \omega $ production.
\vskip 0.3cm

The most interesting quantity is the $\phi /\omega $ ratio in Fig. 5, as here we expect a reduced 
sensitivity to several details of the parametrization (i.~e. less influence from initial and final state
interactions or from the modelling of $p, \omega $ and $\phi $). Then $R_{\phi /\omega} $ decreases with decreasing $Q $ as shown from the ANKE data,
however, there is still a significant variation for different parametrizations within currently accepted limits.
Derivations from ideal $\phi/\omega $ mixing enhance $R_{\phi/\omega}$ by less than 10 \, \%.
\vskip 0.3cm
The conclusions to be drawn are evident. From the experiment side more detailed information on the 
$R_{\phi/\omega}(Q) $ dependence is necessary, to restrict the various model parameters (here information on 
$R_{\phi /\omega} $ at $Q $ very close to threshold would be very interesting, as different models 
predict a significant variation when approaching the $\phi $ and $\omega $ thresholds); in addition, 
angular distributions for both vector mesons, beyond the existing data 
are urgently needed (for example, the $\omega $ and $\phi $ angular distribution at $Q \sim $ 90 MeV 
are fairly isotropic \cite{disto} - \cite{schul}, in striking contrast to the $\omega $-data at Q = 173 MeV 
\cite{abd}, which exhibit a very strong non isotropic structure).
\vskip 0.3cm
From theory, both meson-exchange and quark-gluon models should be tested consistently on existing and
future $\omega, \phi $ data (also to explore the dominant reaction mechanisms at much larger excess 
energies). In addition to still missing ingredients (such as the vector-meson-proton final state 
interaction, relativistic corrections to the production operator and its extension to gluon exchange of next order, or an improvement of $pp $ initial 
state interactions \cite{hanhart,haidenb} most urgent seems a more refined modelling of the 
$q$-diquark structure of the proton (bridging the extremes from a point-like diquark to its resolution as 
two uncorrelated quarks \cite{carl} together with the incorporation of genuine relativistic corrections, 
such as the Lorentz-quenching of the protons in the initial state \cite{amgar, dillig}; 
and a more systematic extension of the formalism to explore the 
role of intermediate baryon resonances both for the $\omega $ and $\phi $ channel \cite{tsush, fuchs}.
Progress along those lines would pave the way to other, even more subtle questions, such as the 
$s \bar{s} $ content of the proton and its influence on its spin structure (\cite{oh, wied}). 
\vskip 2.0cm

\begin{figure}[h!]
\begin{center}
\includegraphics[width=12.0cm,height=10.0cm,angle=0]{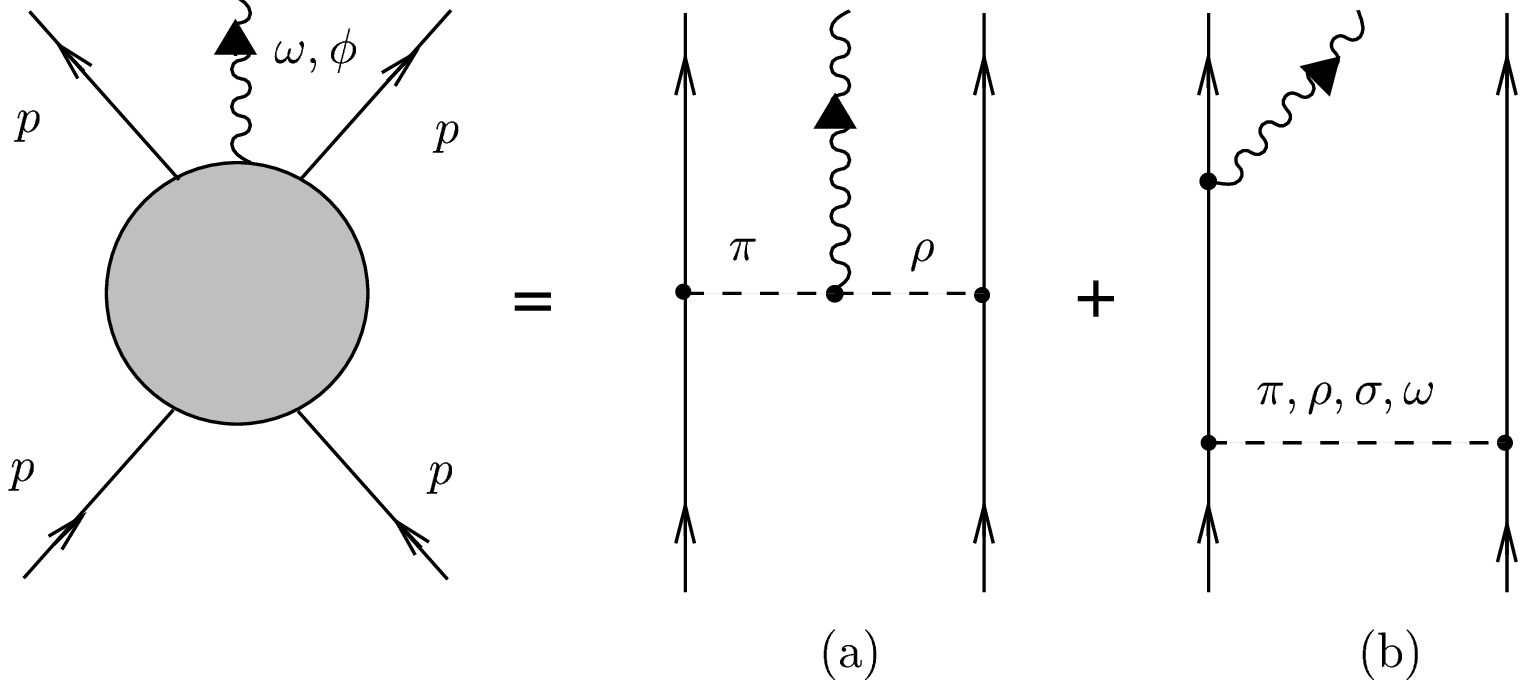}
\caption{Leading contributions to near-threshold $pp \to pp \omega, \phi$ production in meson-exchange models.
(a) 'mesonic' and (b) 'nucleonic' contribution.}
\end{center}
\end{figure}

\begin{figure}[h!]
\begin{center}
\includegraphics[width=12.0cm,height=20.0cm,angle=0]{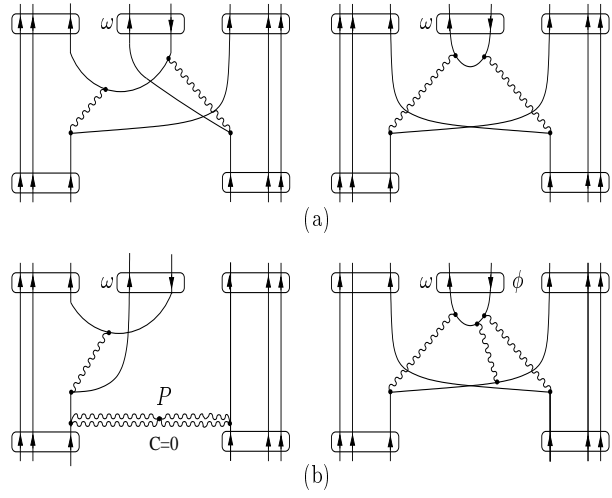}
\caption{As in Fig.~1 however, in a quark-gluon exchange model. (a) 'mesonic' 2-gluon exchange in $\omega$ 
and (b) 'nucleonic' Pomeron-gluon and 3-gluon exchange in $\omega, \phi$ production (for ideal $\omega, \phi $ 
mixing the 'pomeron' component is absent in $\phi $-production).          }
\end{center}
\end{figure}

\begin{figure}[h!]
\begin{center}
\includegraphics[width=12.0cm,height=10.0cm,angle=0]{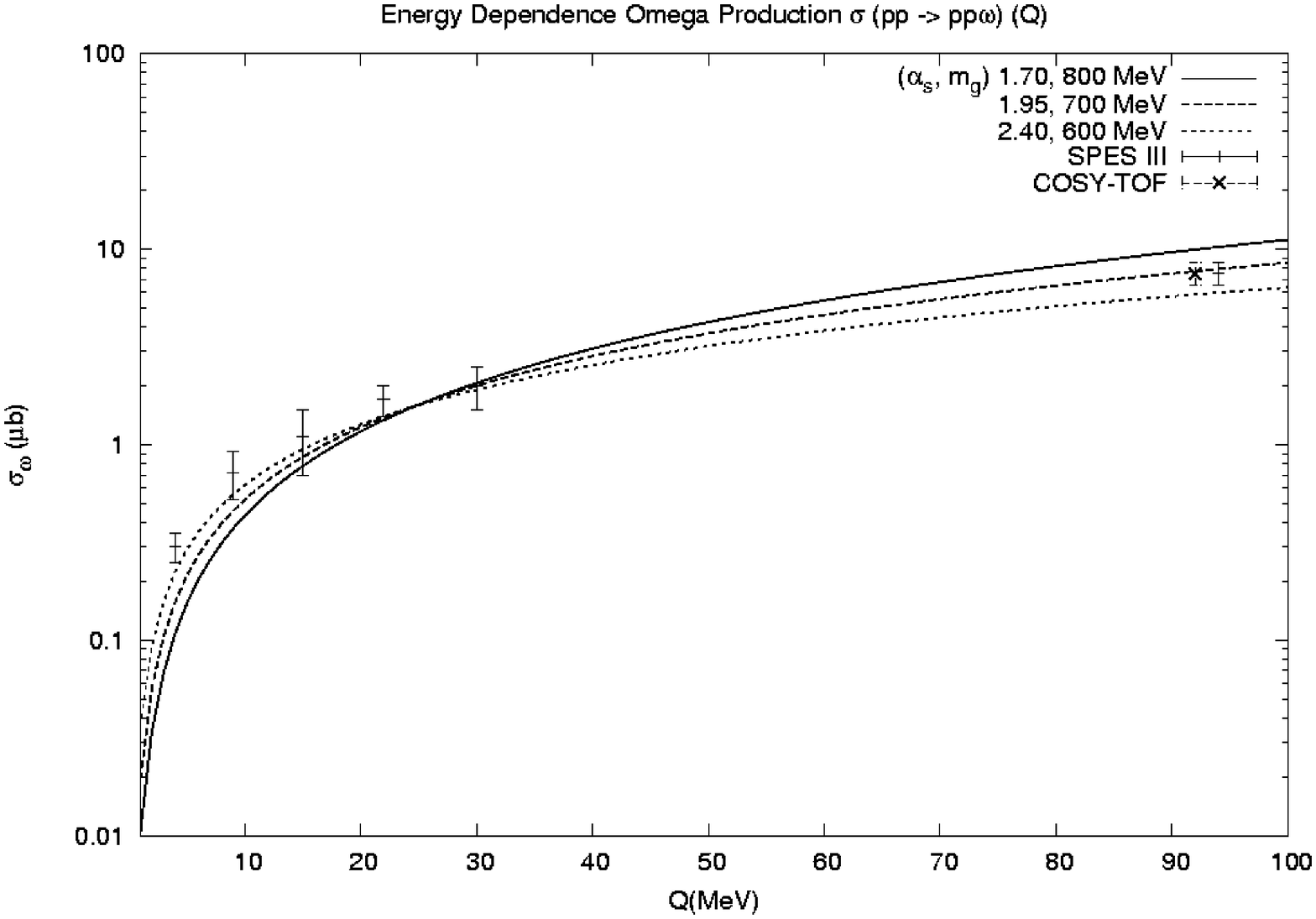}
\caption{Energy dependence of the $pp \to pp \omega$ cross section. Compared are different parametrizations (see legend) with data for $\omega$ production 
\cite{schul} - \cite{hibou}}.
\end{center}
\end{figure}

\begin{figure}[h!]
\begin{center}
\includegraphics[width=12.0cm,height=10.0cm,angle=0]{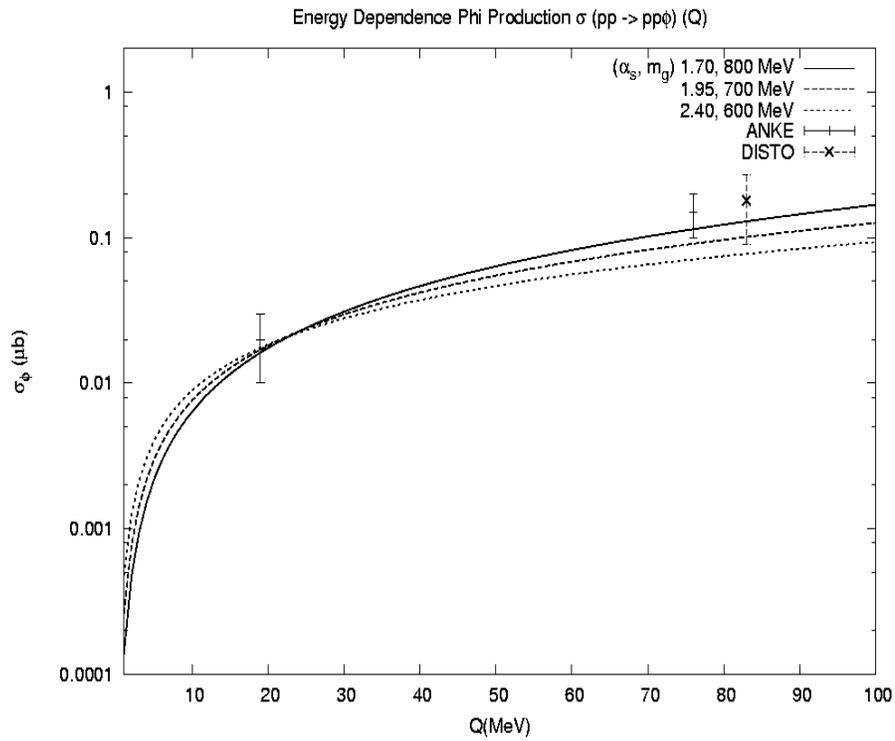}
\caption{As Fig. 3, however for the $pp \to pp \phi $
cross section, compared with data from refs. (\cite{disto, hartm})}.
\end{center}
\end{figure}

\begin{figure}[h!]
\begin{center}
\includegraphics[width=12.0cm,height=10.0cm,angle=0]{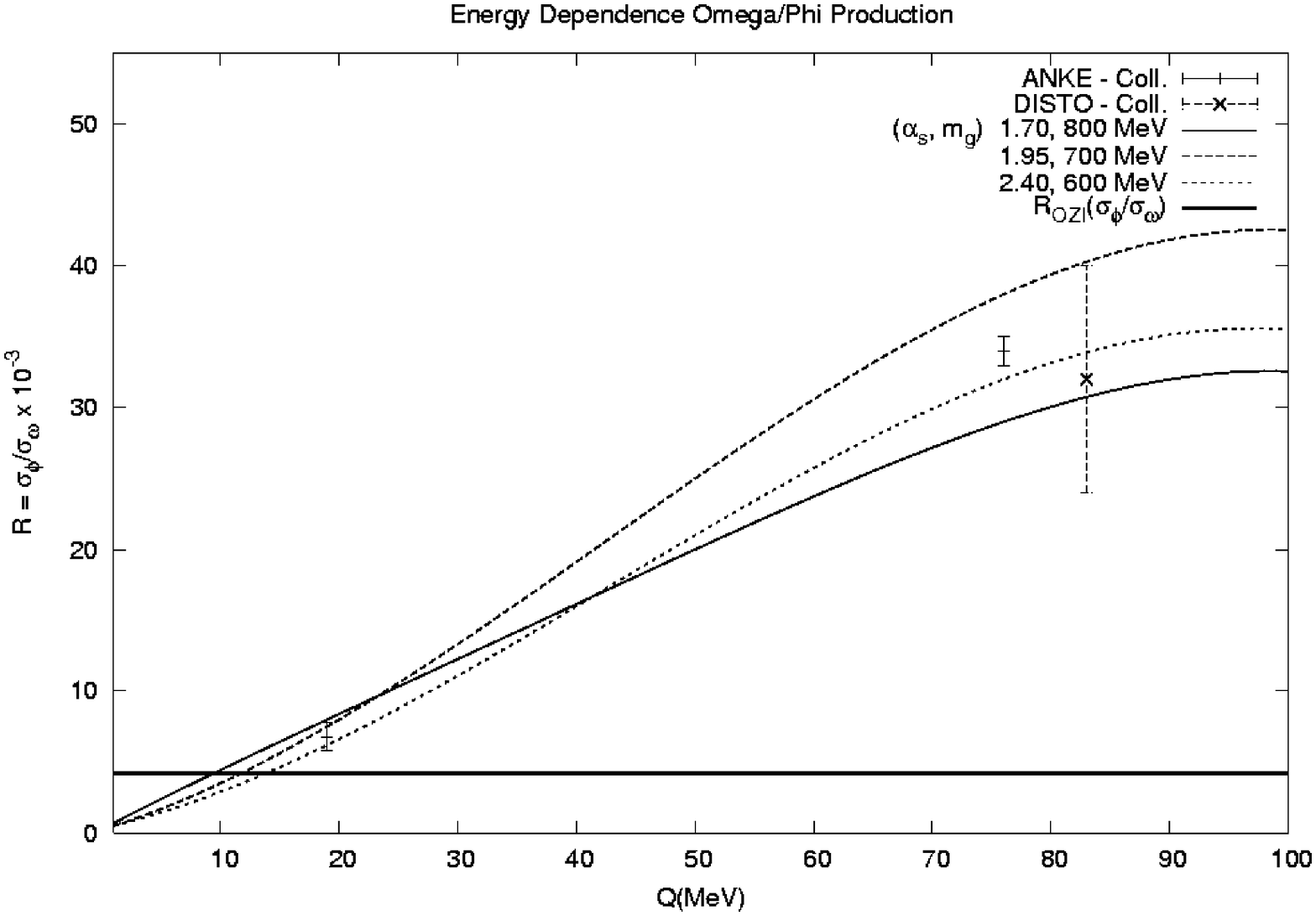}
\caption{Q-dependence of $R_{\phi/\omega} $ for different parametizations, compared to COSY-TOF, ANKE  
and DISTO  data (\cite{disto} -  \cite{abd}).  $R_{\text{OZI}} = 4.2 \cdot 10^{-3} $ shows the OZI 
prediction for nonideal $\omega /\phi$ mixing (see text)}.
\end{center}
\end{figure}

\end{document}